\documentclass[conference]{IEEEtran}
\IEEEoverridecommandlockouts
\usepackage{cite}
\usepackage{amsmath,amssymb,amsfonts}
\usepackage{graphicx}
\usepackage{textcomp}
\usepackage{xcolor}
\usepackage{booktabs}
\usepackage{seqsplit}
\usepackage{tikz}
\usetikzlibrary{shapes,arrows,positioning}
\usepackage[hidelinks]{hyperref}
\def\BibTeX{{\rm B\kern-.05em{\sc i\kern-.025em b}\kern-.08em
    T\kern-.1667em\lower.7ex\hbox{E}\kern-.125emX}}
\begin{document}

\title{EVM-Anchored Provenance for AI-Assisted Code Evolution:\\A Smoke-Scale Deployment Report}

\author{
\IEEEauthorblockN{Einar C\'esar Santos}
\IEEEauthorblockA{\textit{Universidade Federal de Catal\~ao}\\
einar@ufcat.edu.br}
}

\maketitle

\begin{abstract}
We report a smoke-scale deployment of an EVM-anchored provenance framework for the state stream produced by AI-assisted code generation. The system comprises \emph{codebase-state-manager}, an open-source Model Context Protocol (MCP) server (MIT, v0.2.1) that exposes codebase state, transition, and reward operations to any MCP-compatible assistant, and \textsc{CodebaseStateRegistry}, a Solidity contract deployed on Base Sepolia that emits per-state commitment events. Payloads are pinned to the InterPlanetary File System (IPFS) and indexed through a rebuildable SQLite mirror that resynchronizes by replaying contract events. We delimit precisely what anchoring does and does not attest (existence, ordering, and non-equivocation by the committer key; never truthfulness or fabrication-prevention) and outline the orthogonal supply-chain controls required to close that gap. Under an author-produced synthetic workload on a local Anvil simulator, the median commit costs 103{,}282 gas, modeled at US\$0.0018--0.0035 in rollup fees per commit (US\$17.9--34.8 per 10{,}000 states), and simulator latency for the \texttt{create} operation is 539.8~ms ($p_{50}$) and 725.0~ms ($p_{95}$) under loopback storage and instant mining. The live deployment holds 14 \texttt{StateCommitted} events verifiable on-chain; the last consumed 98{,}977 gas at 1\,gwei (about US\$0.30). All empirical figures are descriptive quantiles from a single author run; no testnet latency is reported, and no claim of superiority is made against any baseline. A receipts-trie consumption mechanism, a \texttt{BountyEscrow} use case, and a FAIR-aligned dataset publication pattern are specified as outlook and are not implemented. A property-by-property comparison against OpenTimestamps, Sigstore Rekor, Centrifuge AnchorRegistry, in-toto/SLSA, Radicle, and Lalli~\emph{et~al.} places this work relative to the surrounding provenance landscape.
\end{abstract}


\section{Introduction}
\label{sec:intro}
Coding assistants and autonomous software agents are now embedded in daily development workflows. Each interaction produces a transition in the underlying codebase: a prompt maps to an edit, the edit is applied, the new state is validated, a reward may be computed. At scale, this \texttt{(prompt, diff, reward)} stream is a substrate for auditing and, prospectively, training workflows. The Model Context Protocol (MCP) standardizes how assistants discover and invoke external tools~\cite{anthropic2024mcp}. The first artifact of this paper, \emph{codebase-state-manager}~\cite{santos2026csm}, is an open-source MCP server exposing codebase state management (states, transitions, rewards, and compact LLM-oriented context) to any MCP-compatible assistant, with pluggable persistence backends.

Persistence alone does not serve all audit interests. An operator-controlled database yields fast queries, but parties that do not trust the operator cannot independently verify it. Anchoring entire states on a public blockchain would solve verifiability, yet on-chain payload storage scales poorly: small diffs sit near the economic boundary, while larger states become prohibitive on L1 and L2 alike (quantified in Section~\ref{sec:cost}). The second artifact, and the focus of the evaluation, is an EVM-anchored commitment layer that resolves this trade-off by emitting only canonical commitments on-chain, delegating payloads to IPFS, and indexing queries through a local mirror. It is integrated into codebase-state-manager as an additional persistence mode (\texttt{db\_mode=evm}); for the supported lifecycle subset the assistant-facing tool surface is unchanged, and operations outside that subset raise a documented \texttt{NotImplementedError} (Section~\ref{sec:impl-evm}).

We delimit the guarantee purchased. Anchoring attests \emph{existence at commit time}, \emph{ordering}, and \emph{non-equivocation by the committer key}: post-hoc rewriting, back-dating, and silent deletion become detectable by any third party. It does not attest truthfulness of the committed content; an operator can fabricate tuples before committing them. Section~\ref{sec:threat} delimits the answerable audit questions and the orthogonal supply-chain controls required beyond them.

\subsection*{Scope and contributions of this paper}
\label{sec:scope}
This paper reports an \emph{engineering deployment} at smoke scale. The contributions are two and are stated without inflation: (i)~the design and implementation of the MCP server and its commitment layer, deployed on the Base Sepolia testnet with a public artifact (Sections~\ref{sec:design}--\ref{sec:impl}); and (ii)~a property-level comparison against the surrounding provenance landscape and a single-author, single-run characterization of the deployed layer on a disclosed synthetic workload under one reproducible OP-stack fee model, with explicit threats to validity (Section~\ref{sec:eval}). The contributions are deliberately bounded: no production latency is reported (n=50 simulator states, n=14 testnet commits), no independent re-run is included, and three components---the receipts-trie consumption relayer, a \texttt{BountyEscrow} contract, and a \texttt{DatasetPublisher} contract---are specified as outlook (Section~\ref{sec:roadmap}) and are not implemented. We do not claim novelty of any single primitive in the system; we claim a specific \emph{combination} (assistant-native MCP capture plus per-state EVM commitments on a public rollup, with a specified same-chain consumption path) that, within the surveyed literature, is not operated elsewhere. Head-to-head quantitative comparison with any baseline is future work.

\section{Related Work}
\label{sec:related}

\subsection{Verifiable Anchoring and Off-Chain Storage}
\label{sec:related-anchoring}
OpenTimestamps~\cite{todd2016ots} anchors arbitrary digests into Bitcoin via aggregated Merkle calendars since 2016; it timestamps but emits no EVM-consumable events and models no development-state transitions. Centrifuge's AnchorRegistry~\cite{centrifuge2019anchor} anchors artifact hashes on Ethereum with lineage, targeting business documents. Lamichhane~\emph{et~al.}~\cite{lamichhane2024vdics} pursue verifiable permanence for IPFS clusters via W3C DIDs and Verifiable Credentials. Hasan~\emph{et~al.}~\cite{hasan2023electronics} store CCTV footage on private Ethereum with event-listener search front-ends. Lalli~\emph{et~al.}~\cite{lalli2024ijicr} measure commit costs on Ethereum Layer-1 for academic certificates ($\approx$0.009 ETH each) and report IPFS retrieval $\sim$3$\times$ slower than HTTPS, gateway single points of failure, and unprevented duplicate issuance as open problems. Status against each open problem: retrieval speed remains \emph{open} (our fallback reads are loopback-only); the gateway single point of failure is mitigated by design (any IPLD-compatible pinning service is substitutable) but \emph{not demonstrated} at production scale; duplicate issuance is \emph{deliberately permitted}, with on-chain duplicate \emph{detection} instead of prevention (Section~\ref{sec:dedup}).

\subsection{Supply-Chain and AI Provenance}
\label{sec:related-supply}
The software supply-chain community has standardized provenance attestation without blockchains. in-toto~\cite{torres2019intoto} records signed, inspectable metadata about each step of a software supply chain; SLSA~\cite{slsa} builds provenance levels on top of it; and Sigstore Rekor~\cite{newman2022sigstore} adds a free, publicly verifiable transparency log. These systems answer \emph{who performed which build or release step on which inputs}; they do not model the fine-grained state stream produced \emph{during} AI-assisted development, nor emit events consumable by smart contracts. Our stream sits upstream of them: in-toto-style attestations over our states are the orthogonal control that would cover the fabrication-before-commit gap of Section~\ref{sec:threat}. Radicle~\cite{radicle} provides peer-to-peer Git replication under cryptographic identities with no central controller.

For AI artifacts, L\"uthi~\emph{et~al.}~\cite{luthi2020dlt} implement a graph-based provenance model for AI assets (datasets, models) as a smart contract on a permissionless blockchain, targeting value-chain exchange rather than development events. Wang~\emph{et~al.}~\cite{wang2025ibis} record dataset licensing and lineage for AI training on Daml-on-Canton. Reilly~\emph{Jr.}~\cite{reilly2025rsp} standardizes the Sentinel Evidence Package with payload digests and blockchain timestamp anchors. Tomlinson~\cite{tomlinson2026lockb0x} defines the \emph{Codex Entry} abstraction without specifying how downstream contracts consume commitment events. Velasco~\emph{et~al.}~\cite{velasco2026icsme} study automated, explainable provenance of AI-generated code without blockchain anchoring; their focus on explanation complements our focus on durable, third-party-verifiable recording. The relayer pattern of Section~\ref{sec:consumption} descends from authenticated data-feed work such as Town Crier~\cite{zhang2016towncrier}.

\subsection{Property-Level Positioning}
\label{sec:positioning}
A scoped literature search (with snowballing per Wohlin~\cite{wohlin2014snowballing}, single round, conducted 2026-Q2) grounds this positioning; one snowballing round does not saturate this design space, and the snapshot pre-dates some 2026 work on tamper-evident audit logs for MCP tool calls that we discuss as a gap in Section~\ref{sec:threats}. Table~\ref{tab:position} places our deployment against the five operated alternatives that bound the design space on three properties the literature flags as load-bearing for provenance systems: (a)~\emph{contract-consumable on-chain surface}---does the system emit events that downstream smart contracts can react to without off-chain trust; (b)~\emph{per-commit on-chain cost}---a per-state economic anchor; and (c)~\emph{deployment maturity}---whether the system has been operated with a non-author trace. The table is descriptive and makes no performance claim; columns are populated from each system's published artifacts.

\begin{table}[!t]
\caption{Operated provenance systems by five properties. \emph{Non-author trace} = a deployment not produced by the original authors; \emph{AI workflow} = native support for AI-assisted development events. None of the listed systems has a public head-to-head benchmark against any other.}
\label{tab:position}
\centering
\resizebox{\columnwidth}{!}{%
\scriptsize
\setlength{\tabcolsep}{2pt}
\begin{tabular}{lcccc}
\toprule
\textbf{System} & \textbf{EVM events} & \textbf{Per-commit cost} & \textbf{AI workflow} & \textbf{Non-author trace} \\
\midrule
OpenTimestamps        & No  & free        & No  & Yes \\
Sigstore Rekor        & No  & free        & No  & Yes \\
AnchorRegistry        & Yes & L1 gas      & No  & Yes \\
in-toto / SLSA        & No  & free        & No  & Yes \\
Radicle               & No  & free        & No  & Yes \\
Lalli~\emph{et~al.}   & Yes & $\sim$0.009 ETH (L1) & No & No \\
\textbf{This work}    & Yes & $\sim$\$0.002 (L2) & Yes & No \\
\bottomrule
\end{tabular}%
}
\end{table}

The residual differentiator, stated without inflation: ours is the only surveyed system whose per-state commitments are EVM events on a public Layer-2 rollup, attached to an assistant-native capture surface (MCP), with a specified (not yet implemented) same-chain consumption path. For timestamping alone, Rekor or OpenTimestamps are strictly better on cost and maturity; we recommend them for that use case. We claim no ``first MCP server''.

\section{System Design}
\label{sec:design}

\subsection{Overview and Threat Model}
\label{sec:threat}
The codebase-state-manager~\cite{santos2026csm} structures the interaction between a coding assistant and the persistence layer for code evolution. Its domain model comprises \emph{States} (numbered snapshots recording prompt, diff, content hash, timestamp), \emph{Transitions} (edges carrying the action and optional reward), and metadata. Thin MCP wrappers call a service layer over pluggable repository backends (SQLite, Neo4j, EVM), failing fast on misconfiguration.

Four adversary classes. (i)~\emph{The backend operator} may back-date, misattribute authorship, or fabricate tuples before committing; anchoring defeats the first two, not the third. (ii)~\emph{The IPFS provider} may serve stale or no payload; pin to $N \geq 3$ services. (iii)~\emph{The centralized L2 sequencer} may censor without chain rewrite; we assume eventual liveness, with L1 forced inclusion as the recovery path at L1 delay and cost. (iv)~\emph{Key compromise} enables scalable fake provenance; mitigation is a threshold multisig or a committer key held in an HSM/KMS, with rotation through \texttt{RoleGranted}/\texttt{RoleRevoked} (Section~\ref{sec:impl-contract}). Three scope decisions. First, confidentiality: public-by-default targets open research and educational artifacts; proprietary codebases require client-side encryption or a private swarm. Second, publication hygiene: payloads are immutably public once pinned, so pre-publish secret-scanning (gitleaks-class) is a required production control; the PoC, operating on author-generated states, does not bundle one. Third, honest Layer-2 consensus with high probability.

\textbf{What the anchored stream answers.} Because fabrication-before-commit is outside the guarantee, we state which audit questions a third party \emph{can} answer: (Q1)~\emph{Did the operator rewrite, reorder, or delete history after the fact?} Detectable: hashes and the monotonic counter are public and append-only. (Q2)~\emph{Did a claimed state exist at a specific time?} Yes, bounded by block timestamp. (Q3)~\emph{Which key committed it, and is a retrieved payload the committed content?} Yes: non-repudiable up to key compromise, verifiable via on-chain hash and JCS canonicalization. What the stream \emph{cannot} answer (prompt genuineness, reward correctness, build success) requires controls orthogonal to anchoring: signed capture at the source, in-toto-style attestations~\cite{torres2019intoto}, or on-chain predicate recomputation. Anchoring converts undetectable tampering into detectable fabrication; we present that conversion, not truth, as the guarantee.

Under these assumptions: \textbf{G1} (Integrity): any third party can verify that a state existed when the chain committed it (Q1--Q3); \textbf{G2} (Public Verifiability): commitments and CIDs are public without operator cooperation; \textbf{G3} (Composable Evidence, design goal): commitments are consumable by downstream systems, off-chain without restriction and, once the mechanism of Section~\ref{sec:consumption} is implemented, by same-chain contracts within its stated window. Contracts have no log-reading opcode, so G3 is a claim about that specified mechanism, not a native capability.

\subsection{Architecture}
\label{sec:arch}
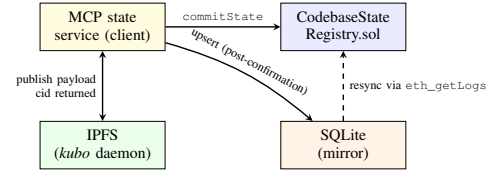
\begin{figure}[!t]
\centering
\resizebox{0.72\columnwidth}{!}{%
\begin{tikzpicture}[
  box/.style={draw, rectangle, minimum width=2.5cm, minimum height=0.95cm, align=center, font=\small, inner sep=3pt},
  evm/.style={box, fill=blue!8},
  ipfs/.style={box, fill=green!8},
  sql/.style={box, fill=orange!10},
  cli/.style={box, fill=yellow!15},
  arrow/.style={->, thick, >=stealth},
  node distance=1.4cm and 2.3cm
]
\node[cli] (client) {MCP state\\service (client)};
\node[evm, right=of client] (reg) {CodebaseState\\Registry.sol};
\node[ipfs, below=of client] (ipfsn) {IPFS\\(\textit{kubo} daemon)};
\node[sql, below=of reg] (sqln) {SQLite\\(mirror)};

\draw[arrow] (client) -- node[above,font=\scriptsize]{\texttt{commitState}} (reg);
\draw[<->, thick, >=stealth] (client) -- node[left,font=\scriptsize,align=right]{publish payload\\cid returned} (ipfsn);
\draw[arrow] (client) to[bend left=12] node[midway,above,sloped,font=\scriptsize]{upsert (post-confirmation)} (sqln);
\draw[arrow,dashed] (sqln) -- node[right,font=\scriptsize]{resync via \texttt{eth\_getLogs}} (reg);
\end{tikzpicture}}%
\caption{Architecture and write path. The client publishes the payload to IPFS, submits \texttt{(stateNumber, stateHash, cid)} to the contract, and upserts the mirror after confirmation; the mirror resynchronizes by replaying contract events.}
\label{fig:arch}
\end{figure}

Figure~\ref{fig:arch} depicts the three-layer architecture. The on-chain \textsc{CodebaseStateRegistry} holds only minimal state (total-state counter, latest-state index, per-state payload hash, and IPFS CID) and emits events on change. Payloads live on IPFS via a local \texttt{kubo} daemon (any IPLD-compatible pinning service is substitutable). A local SQLite mirror records each tuple after confirmation and serves read-side queries that are economically inappropriate on-chain. The MCP state service is the sole client of all three layers; it publishes the payload first and only then submits the resulting CID. 

\subsection{Canonicalization and Deduplication}
\label{sec:dedup}
Every state is canonicalized through JCS (RFC~8785~\cite{rfc8785}) and hashed with \texttt{keccak256}; the digest is stored on-chain as \texttt{stateHash}. The contract keys entries by a strictly increasing \texttt{stateNumber} and reverts on duplicate submission. Canonicalization makes duplicates \emph{detectable}: identical payloads yield identical hashes. Uniqueness is deliberately not enforced: re-committing identical content is sometimes legitimate (e.g., reverting to a prior state). We claim duplicate detection and state-number uniqueness, not impossibility of duplicate registration.

\subsection{Data Flow and Mirror Resynchronization}
\label{sec:dataflow}
\texttt{commitState} is the sole write path: (i)~the client canonicalizes the state; (ii)~uploads the payload to IPFS and obtains a CID; (iii)~submits \texttt{(stateNumber, stateHash, cid)} to the contract; (iv)~the contract emits \texttt{StateCommitted}; (v)~after confirmation, the client upserts the mirror. Step (v) opens a divergence window: a crash between confirmation and upsert leaves the mirror behind. The mirror resynchronizes explicitly: it compares \texttt{totalStates()} with its maximum \texttt{state\_number} and replays \texttt{StateCommitted} events via \texttt{eth\_getLogs} from the last mirrored block, with idempotent upserts. Replay is bounded to blocks at least twelve confirmations deep to tolerate shallow reorganizations. On Base, reorgs are sequencer-level events, not miner races: twelve blocks ($\approx$24\,s) cover shallow sequencer reordering and short L1 reorgs propagating through derivation; deep L1 reorgs are exceptional recovery scenarios. The chain is the source of truth; the mirror is a rebuildable cache: misses fall back to a chain view plus an IPFS fetch.

\subsection{Specified Consumption Path (Not Implemented)}
\label{sec:consumption}
The contract emits six events (\texttt{StateCommitted}, \texttt{TransitionCommitted}, \texttt{RewardUpdated}, \texttt{StateRevoked}, \texttt{RoleGranted}, \texttt{RoleRevoked}). EVM logs are write-only: no opcode lets a contract read another contract's logs. We specify the standard explicit mechanism: a permissionless relayer observes the event off-chain and submits to a downstream contract the event data, the block header, and a Merkle--Patricia proof of the log's inclusion in that block's receipts trie, verified against the \texttt{receiptsRoot} reachable via \texttt{BLOCKHASH} for the most recent 256 blocks. \textbf{This path is specified but not implemented}: no relayer, verifier contract, or gas measurement exists in the deployed artifact, and every statement about it here concerns the specification, not a deployed mechanism. Its cost is estimable: a receipts-trie inclusion proof carries $\sim$6--10 RLP nodes ($\approx$1--2\,kB calldata) plus per-node hashing and RLP decoding, bounding verification at $10^{5}$--$10^{6}$ gas ($\approx$US\$$0.002$--$0.02$ at Table~\ref{tab:cost} prices)---a fee-model estimate, not a measurement. Three scope limits bound every downstream claim here. (a)~\emph{Window}: $\approx$8.5 minutes on Base ($\sim$2\,s blocks); older claims require optimistic attestation with a challenge period, reintroducing a liveness-on-watchers assumption we state openly. (b)~\emph{Same-chain} only. (c)~\emph{Batching}: under Merkle batching (Section~\ref{sec:cost}) per-state events cease to exist, so receipts-trie proofs do not apply; batched states would need membership proofs against the anchored root, not specified here. G3 therefore covers only the per-state, event-emitting path. The per-state mode is deployed because per-state events are the substrate G3 consumes: batching is $\sim$$200\times$ cheaper but forfeits that path, and at US\$$17.9$--$34.8$ per 10{,}000 states the premium fits audit budgets. Per-state anchoring is the default for G3-dependent deployments; batching serves anchoring-only ones.

\section{Implementation}
\label{sec:impl}

\subsection{MCP Server}
\label{sec:impl-mcp}
\emph{Codebase-state-manager} is implemented in Python on FastMCP and runs as a local stdio subprocess under any MCP-compatible assistant, exposing 26 tools in four families (lifecycle and state capture, history inspection, compact context and rewards, runtime recovery). A compact context representation, SCC-E (State Compression Code, Embedding), stores LLM-oriented state summaries so long sessions reconstruct context without replaying full history. Compact queries attach the reward of the earliest producing transition, coupling state, action trail, and reward for future reinforcement-learning workflows; no RL experiment is claimed. A repository layer hides the backend (SQLite, Neo4j, EVM) behind one interface. The test suite spans 568 tests (unit through stress layers), with \texttt{mypy}/\texttt{bandit} clean. Test counts measure activity, not adequacy: we report no coverage or mutation figures, and treat that as a limitation.

\subsection{Smart Contract and Revocation Semantics}
\label{sec:impl-contract}
\textsc{CodebaseStateRegistry} is written in Solidity and built, tested, and gas-profiled with Foundry~\cite{bookfoundry} (\texttt{optimizer\_runs=200}, the conventional setting for contracts called at moderate frequency). The interface is the minimal subset required by the repository layer: \texttt{commitState}, \texttt{getStateHash}, \texttt{getStateCid}, \texttt{getCurrentState}, \texttt{totalStates()}, and \texttt{revokeState(uint64)}. Revocation sets a flag and emits \texttt{StateRevoked}; it \emph{preserves} \texttt{stateHash} and CID, so the existence attestation (G1) survives revocation while consumers are signaled that the state is no longer operative; \texttt{stateNumber} is never reused; the mirror applies a tombstone on replay.

\textbf{Authorization model.} A \texttt{DEFAULT\_ADMIN\_ROLE} grants and revokes the \texttt{COMMITTER\_ROLE} that alone may call \texttt{commitState} and \texttt{revokeState}; grants and revocations emit the auditable \texttt{RoleGranted}/\texttt{RoleRevoked} events, making rotation public and Q3 (\emph{which key committed it}) evaluable against on-chain senders. The Base Sepolia deployment holds both roles on a single author key---acceptable at smoke scale but the weakest live link; production configuration is a threshold multisig admin with an HSM/KMS-backed committer and scheduled rotation (Section~\ref{sec:threat}). Seven Foundry tests cover happy paths, duplicate rejection, unauthorized access, and basic revocation; extended revocation edge cases are queued, so revocation is specified and partially tested, not fully verified.

\subsection{EVM Backend Components}
\label{sec:impl-evm}
Within the repository layer, \texttt{repositories/evm/} adds the fourth persistence mode. The Python canonical encoder implements RFC~8785~\cite{rfc8785} over dictionaries and lists, using \texttt{pycryptodome} for \texttt{keccak256}; it is the single authority on which byte sequence is hashed for any state. Conformance relies on UTF-8 round-trip fixtures; JCS edge cases (number serialization, I-JSON) are not yet exercised against the RFC's reference vectors (queued task). Since the encoder is the single hashing authority and ships with the artifact, verifiers reproduce our hashes exactly; the risk is divergence from \emph{other} JCS implementations. The IPFS client targets a local \texttt{kubo} daemon over stdlib HTTP with a localhost allow-list. The contract client uses \texttt{web3.py}, the standard Python library for interacting with EVM nodes over JSON-RPC, with EIP-1559 envelopes. \texttt{EVMStateRepository} implements the \texttt{create}, \texttt{get\_by\_number}, \texttt{get\_current}, and \texttt{create\_next} subset of the existing interface; other operations raise a documented \texttt{NotImplementedError}. Misconfiguration fails fast rather than silently downgrading to SQLite; an unreachable RPC aborts startup, and a commit-time failure surfaces as a tool error without updating the mirror.

\subsection{Tests, Deployment, and Artifact}
\label{sec:deployment}
Seventy-five unit tests (15 encoder, 11 IPFS, 11 contract client, 16 repository, 9 measurement, 3 Anvil, 2 parity) plus the 7 Foundry tests pass. Fuzz and invariant campaigns are future work. The contract is live on Base Sepolia at \texttt{\seqsplit{0xA326f4A75F00189aa6fCec71289D2c05Bc3D5310}} with 14 \texttt{StateCommitted} events; deploy transaction \texttt{\seqsplit{0x4e0fdc05b1200a38c39996a2ee504be78db472563848c36759578eb24d852fb5}}; representative commit (state 13) \texttt{\seqsplit{0xe5c855e8b06a32a902d78e364ac1985e248a881e1afe9c0a45a2f980963045eb}}. These events are \emph{publicly} verifiable by any third party via independent RPC calls; we do not claim independent verification by others. The deployer was funded from the Coinbase Developer Platform faucet~\cite{coinbase2024base} (no affiliation). The full artifact (server, EVM backend, contract, tests, and the scripts generating every table, including the fee model) is public~\cite{santos2026csm}.

\section{Evaluation}
\label{sec:eval}

\subsection{Methodology}
\textbf{S1} (local): 50 synthetic states (35 small ($\sim$2\,kB, the empirical median diff size), 12 medium ($\sim$116\,kB), 3 large ($\sim$780\,kB)) driven against a local Anvil instance (solc 0.8.24) with instant mining, on a Linux x86-64 host (8-core, 32\,GB RAM; Python 3.12; \texttt{kubo}). We report end-to-end latency for \texttt{create}, mirror read, and fallback read, and gas from \texttt{forge test -{}-gas-report}. The Table~\ref{tab:lat} thresholds are \emph{usability thresholds}, fixed before measurement (interactive mirror reads; batch-tolerant \texttt{create}); they are not performance targets, and with medians $3$--$5.5\times$ below them the informative content is the distribution, not the pass.  \textbf{S2} (testnet): 14 commits against the deployment of Section~\ref{sec:deployment}, verified through independent RPC calls. All measurements are author-produced; S1 is synthetic and simulator-based. S2 measured no latency (RPC rate limits and faucet budget capped it at 14 commits); an $n \geq 100$ testnet latency campaign precedes any production latency claim. A direct head-to-head measurement against Sigstore Rekor, OpenTimestamps, or any other baseline is future work; we report no such comparison in this paper.

\subsection{Latency}
\label{sec:latency}
Table~\ref{tab:lat} reports $p_{50}$/$p_{95}$ \emph{for S1 only}. These figures capture client- and storage-side overhead, not confirmation, remote pinning, or sequencer behavior; no system-level latency is claimed. With $n=50$, $p_{95}$ is a 2--3-sample order statistic; quantiles are descriptive, without confidence intervals. \texttt{create} is dominated by EIP-1559 submission and the IPFS \texttt{add}; the fallback read by the loopback gateway round-trip.

\begin{table}[!t]
\caption{S1 (Anvil simulator) latency vs.\ usability thresholds; descriptive quantiles, single author run.}
\label{tab:lat}\centering
\scriptsize
\setlength{\tabcolsep}{4pt}
\begin{tabular}{lrrrr}
\toprule
\textbf{Operation} & \textbf{$n$} & \textbf{$p_{50}$ (ms)} & \textbf{$p_{95}$ (ms)} & \textbf{Threshold (ms)} \\
\midrule
\texttt{create} & 50 & 539.8 & 725.0 & $\leq$ 3000 \\
\texttt{get\_by\_number} (mirror) & 50 & 3.31 & 4.69 & $\leq$ 10 \\
\texttt{get\_by\_number} (fallback) & 50 & 514.7 & 563.7 & $\leq$ 1000 \\
\bottomrule
\end{tabular}
\end{table}
\subsection{Gas, Cost Model, and Storage Break-Even}
\label{sec:cost}
The median commit costs 103{,}282 gas in the isolated-commit micro-benchmark; contract tests span warm (26{,}847) to cold (161{,}196) storage. A fee-schedule derivation predicts 53{,}000--65{,}000 gas; the empirical median is $1.6$--$1.9\times$ higher, attributable to cold-storage access and a \texttt{totalStates++} \texttt{SSTORE} the derivation omitted (a cold zero-to-nonzero \texttt{SSTORE} costs 22{,}100 gas under EIP-2929, consistent with most of the gap). The last testnet commit consumed 98{,}977 gas ($-$4.2\% vs.\ the simulator median), consistent with the cold-to-warm transition expected on a populated contract.

Fee models pricing only L2 execution gas underestimate rollup costs: on OP-stack rollups such as Base, the total fee is $\mathit{L2\_exec} + \mathit{L1\_data}$~\cite{opstack2024fees}, and the L1 data fee can dominate in the low-congestion regime quoted. All monetary values below come from a released script implementing the full Ecotone fee function:
\begin{equation}
\label{eq:fee}
\mathit{L1\_data} = \frac{\mathit{estSize}\,(16\,s_b\,f_b + s_{\beta}\,f_{\beta})}{10^{6}}\,,
\end{equation}
where $\mathit{estSize} = (4 z + 16 n)/16$ over $z$ zero and $n$ nonzero tx bytes; we use Base's posted scalars ($s_b{=}1101$, $s_{\beta}{=}659851$; both governance-updatable) and two parameter points: \emph{low} ($f_b{=}15$\,gwei, $f_{\beta}{=}0.05$\,gwei) and \emph{high} ($f_b{=}100$, $f_{\beta}{=}1.0$), with L2 gas price 0.005\,gwei and ETH at US\$$3{,}000$. For a $\sim$300-byte commit this yields an L1 data fee of US\$$0.0002$--$0.0019$. The model is linear, so a $\pm 50\%$ move in any input moves the corresponding total proportionally. Figures are \emph{modeled}, not measured on mainnet Base, and are lower bounds under L1 fee spikes.

\begin{table}[!t]
\caption{Per-commit gas and modeled fee cost under (\ref{eq:fee}); \emph{Exec} is L2 execution only, \emph{Fee} adds the modeled L1 data fee.}
\label{tab:cost}
\centering
\scriptsize
\setlength{\tabcolsep}{3pt}
\begin{tabular}{lrrrr}
\toprule
\textbf{Strategy} & \textbf{Gas} & \textbf{Exec US\$} & \textbf{Fee US\$} & \textbf{US\$/10k} \\
\midrule
per-state (median) & 103{,}282 & 0.0015 & 0.0018--0.0035 & 17.9--34.8 \\
per-state (warm) & 26{,}847 & 0.0004 & 0.0006--0.0023 & 6.4--23.4 \\
per-state (cold) & 161{,}196 & 0.0024 & 0.0027--0.0044 & 26.6--43.5 \\
batch $N{=}100$ $^{\dagger}$ & 450 & $6.8{\times}10^{-6}$ & $10^{-5}$--$10^{-5}$ & 0.08--0.16 \\
\bottomrule
\multicolumn{5}{l}{\scriptsize $^{\dagger}$ Projected: one 45{,}000-gas anchor over 100 states; batching emits no}\\
\multicolumn{5}{l}{\scriptsize per-state events, forfeiting the Section~\ref{sec:consumption} path. Per-10k: fees $\times$ $10^{4}$.}
\end{tabular}
\end{table}

\textbf{Fee ranges and the batch row.} The low--high points span the modeled L1 data fee. Gas rows are measured; the batch row is projected: the deployed interface has no batch entry point, and adding one mid-study would invalidate the artifact.

\textbf{What the fee figures exclude.} Table~\ref{tab:cost} reports gas and rollup fees only; the full stack adds IPFS pinning (all 50 payloads under 1\,GB; single-digit US\$/month at $N{\geq}3$ redundancy on public pinning services), RPC access (free tiers suffice at PoC scale), and mirror storage---modest at PoC volumes, dominant at high throughput, excluded as provider-dependent.

\textbf{Storage break-even.} Storing the payload itself as calldata costs $21{,}000 + 16$\,gas per nonzero byte. Against the median commit, the gas-only break-even payload is $\approx$5.1\,kB: below it, naive calldata is cheaper in gas but buys no content addressing, deduplication, or uniform retrieval. For the 70\% of the workload below break-even ($\sim$2\,kB diffs), the justification is not raw gas but the verifiable, content-addressed, uniformly queryable stream, at a quantified premium (US\$$0.0018$--$0.0035$ per state) over near-zero transparency-log timestamping (Section~\ref{sec:positioning}). For the 30\% tail the economics are decisive: 116\,kB and 780\,kB states would cost $\approx$1.9\,M and $\approx$12.8\,M gas ($\approx$US\$$0.13$ and $\approx$US\$$0.85$ at the low point), and on Layer~1 $\approx$US\$$2.42$ and $\approx$US\$$576$ at 15\,gwei.

\subsection{Live Testnet Evidence}
The Base Sepolia deployment currently holds 14 commits (states 0--13). The last commit consumed 98{,}977 gas at 1\,gwei, $200\times$ the projection assumption, pricing it at $\approx$US\$$0.30$: evidence of executability and the upper bound of the abstract's modeled range. Testnet prices are unrepresentative of mainnet Base in both directions; the live commit confirms that the contract executes end-to-end under realistic fee conditions, not that the modeled per-commit price is accurate.

\subsection{Threats to Validity}
\label{sec:threats}
(i)~\emph{Self-produced evidence}: author-produced measurements against author-set thresholds; independent re-runs precede any production claim; the artifact is released to enable them. (ii)~\emph{Simulator latency}: S1 excludes confirmation, remote pinning, and sequencer paths. (iii)~\emph{Cost-model risk}: monetary rows are modeled, not measured on mainnet Base; the live testnet commit price was 200$\times$ the modeled L2 gas price. (iv)~\emph{Workload realism}: the 35/12/3 distribution is synthetic; tighter clustering around small payloads weakens per-state economics toward the break-even. (v)~\emph{Statistics}: quantiles are descriptive, single-run, without confidence intervals. (vi)~\emph{Availability and finality}: pin $N\geq3$; high-value workflows should wait the 12-block confirmation window. (vii)~\emph{Assisted-path coverage}: the EVM backend was exercised through the repository layer and measurement scripts, never by an external MCP assistant. All empirical claims here are backend-level (gas, fees, latency, durability) and independent of the driving client, so the gap weakens no reported number; it blocks assistant-level usability claims, which we do not make. (viii)~\emph{Literature snapshot}: the positioning table of Section~\ref{sec:positioning} reflects a single 2026-Q2 snowballing round; tamper-evident audit logging for MCP tool calls has emerged as an adjacent line of work in 2026 that we do not yet position against in depth.

\subsection{Property Comparison Against Operated Baselines}
\label{sec:comparison}
The positioning table of Section~\ref{sec:positioning} summarizes five operated baselines. We restate the residual differentiator without inflation: ours is the only surveyed system whose per-state commitments are EVM events on a public Layer-2 rollup, attached to an assistant-native capture surface (MCP), with a specified (not yet implemented) same-chain consumption path. For timestamping alone, Rekor or OpenTimestamps are strictly better on cost and maturity; we recommend them for that use case. A measured head-to-head comparison on the same workload and hardware is future work and is the principal missing evaluation in this paper.

\section{Outlook}
\label{sec:roadmap}
Three components are specified below and are not implemented in the deployed artifact. We claim no contribution from them beyond the specification itself; each is the principal missing evaluation or implementation in this paper and is the priority next step before any production claim.

\subsection{Programmable Bug Bounties}
A sponsor instantiates \texttt{BountyEscrow.sol}, an ERC-20 escrow locking USDC against a declared bug class and reward predicate. A claimant calls \texttt{claim(stateNumber, rewardData, header, proof)}, with \texttt{proof} a receipts-trie inclusion proof verified against the \texttt{receiptsRoot} of a header inside the 256-block window; the escrow checks the predicate and releases USDC. The window is severe ($\approx$8.5 minutes); the practical path is optimistic claims with a 24\,h challenge period, bonded challenges, and slashing, all specified but not implemented. The chain attests existence and ordering, not reward correctness; distrusting sponsors can require on-chain predicate recomputation or disputes.

\subsection{FAIR-Aligned Dataset Publication}
A curator commits a Merkle root over a window of $N$ rewarded transitions onto a \texttt{DatasetPublisher.sol} contract and publishes the payloads to a public mirror. Consumers query the registry for each \texttt{stateHash} and rebuild the root locally; per-state proofs allow $O(\log N)$ spot-checks. The pattern is aligned with the FAIR principles, not certified; no formal assessment was applied.

\subsection{Deployment Constraints}
Diffs and prompts may embed personal data, and payloads pinned to public IPFS are effectively unerasable, in tension with erasure rights (Brazil's LGPD, Art.~16; EU GDPR, Art.~17). Production use therefore requires data minimization at capture, pre-publish secret-scanning, client-side encryption or a private swarm, and a documented legal basis. The PoC processes only author-generated states and claims no compliance.

\section{Conclusion}
\label{sec:conc}
We presented \emph{codebase-state-manager}, an open-source MCP server for structured provenance of AI-assisted code evolution, and \textsc{CodebaseStateRegistry}, its EVM-anchored commitment layer, live on Base Sepolia. This is an engineering deployment report at smoke scale: the contributions are a working design and deployment, a property-level comparison against the surrounding provenance landscape, and a single-author, single-run characterization of the deployed layer under one reproducible OP-stack fee model. We delimited the guarantee precisely (existence, ordering, non-equivocation; never truthfulness) and positioned the system as complementary to supply-chain frameworks and free transparency logs, which remain the right choice for timestamping alone. Priority future work: (i)~implementation of the receipts-trie relayer with one end-to-end \texttt{BountyEscrow} claim, turning G3 from design goal into a measured property; (ii)~a direct head-to-head measurement against Sigstore Rekor and OpenTimestamps on the same workload and hardware; (iii)~an $n \geq 100$ testnet latency campaign, the only path to defensible production latency claims; (iv)~a non-author trace reproducing the deployed contract on a developer-owned AI-assisted coding session; and (v)~RFC~8785 reference-vector conformance, revocation edge-case tests, and fuzz/invariant campaigns on the contract.

\begingroup
\scriptsize
\setlength{\itemsep}{0pt}

\endgroup

\end{document}